\documentclass{article}
\usepackage[colorlinks=true,linkcolor=blue]{hyperref}%
\usepackage{authblk}
\usepackage{graphicx}
\usepackage{float}
\usepackage{amsmath}
\usepackage{tikz}
\usepackage{amssymb}
\usepackage{esint}
\usepackage{wasysym}
\usepackage{geometry}
\usepackage{comment}
\usepackage{xcolor}
\newcommand{\tumbleRate}{\alpha}
\newcommand{\action}{\AC}
\newcommand{\actionPert}{\action_{\text{\tiny pert}}}
\DeclareMathOperator*{\SumInt}{%
\mathchoice%
  {\ooalign{$\displaystyle\sum$\cr\hidewidth$\displaystyle\int$\hidewidth\cr}}
  {\ooalign{\raisebox{.14\height}{\scalebox{.7}{$\textstyle\sum$}}\cr\hidewidth$\textstyle\int$\hidewidth\cr}}
  {\ooalign{\raisebox{.2\height}{\scalebox{.6}{$\scriptstyle\sum$}}\cr$\scriptstyle\int$\cr}}
  {\ooalign{\raisebox{.2\height}{\scalebox{.6}{$\scriptstyle\sum$}}\cr$\scriptstyle\int$\cr}}
}

\newcommand{\se}{\operatorname{se}}
\newcommand{\ce}{\operatorname{ce}}

\newcommand{\gmatrix}[1]{\underline{\underline{#1}}}
\newcommand{\Amatrix}{\gmatrix{A}}
\newcommand{\Gmatrix}{\gmatrix{G}}
\newcommand{\Tmatrix}{\gmatrix{T}}
\newcommand{\diag}{\operatorname{diag}}
\newcommand{\TC}{\mathcal{T}}
\tikzset{
xxtsubstrate/.style={decorate, 
line width=1pt,
draw=olive, 
decoration=snake, 
segment amplitude=0.75mm, 
line after snake=0.25mm,
line before snake=0.25mm
},
tsubstrate/.style={decorate, 
line width=1pt,
draw=olive, 
decoration=snake, 
segment amplitude=0.5mm, 
segment length=5pt,
segment amplitude=0.2mm, 
line after snake=1mm,
line before snake=1mm
},
Bsubstrate/.style={decorate, 
line width=1pt,
draw=olive, 
decoration=snake,
segment length=5pt,
segment aspect=0,
segment amplitude=0.5mm, 
line after snake=0mm,
line before snake=0mm
},
substrate/.style={decorate, 
line width=1pt,
draw=olive, 
decoration=snake, 
segment length=5pt,
segment amplitude=0.5mm, 
line after snake=0.5mm,
line before snake=0.5mm
},
activity/.style={very thick,draw=red,postaction={decorate},
decoration={markings,mark=at position .5 with
{\arrow[draw=red]{>}}}},
tactivity/.style={thick,draw=red,postaction={decorate},
decoration={markings,mark=at position .5 with
{\arrow[draw=red]{>}}}},
tEPSactivity/.style={thick,draw=red,postaction={decorate},
decoration={markings,mark=at position .55 with
{\arrow[draw=red]{>}}}},
tAactivity/.style={thick,draw=red},
Aactivity/.style={very thick,draw=red},
Bactivity/.style={very thick,draw=blue,dashed},
tSactivity/.style={thick,draw=red,postaction={decorate},
decoration={markings,mark=at position .7 with
{\arrow[draw=red]{>}}}},
Sactivity/.style={very thick,draw=red,postaction={decorate},
decoration={markings,mark=at position .7 with
{\arrow[draw=red]{>}}}}
}
\tikzset{
  terminal/.style = {
    rectangle,
    align = center,
    minimum size = 6mm,
    rounded corners = 3mm,
    very thick,
    draw = black!50,
    top color = white,
    bottom color = black!20,
  }
}
\usepackage{amsmath}
\usepackage{amsfonts}
\usepackage{cases}
\usepackage{xspace}
\usepackage{psfrag}
\usepackage{subfig}
\usepackage{color}
\usepackage{xcolor}

\newcommand{\ave}[1]{\left\langle #1 \right\rangle}
\newcommand{\spave}[1]{\overline{#1}}

\newcommand{\imag}{\mathring{\imath}}

\newcommand{\plaind}{\mathrm{d}}
\newcommand{\dint}[1]{\mathchoice{\!\plaind#1\,}{\!\plaind#1\,}{\!\plaind#1\,}{\!\plaind#1\,}}
\newcommand{\ddint}[1]{\ddintx{#1}{d}}
\newcommand{\ddintx}[2]{\mathchoice{\!\plaind^{#2}#1\,}{\!\plaind^{#2}#1\,}{\!\plaind^{#2}#1\,}{\!\plaind^{#2}#1\,}}
\newcommand{\dTwoint}[1]{\ddintx{#1}{2}}
\newcommand{\ddMOneint}[1]{\mathchoice{\!\plaind^{d-1}#1\,\,}{\!\plaind^{d-1}#1\,\,}{\!\plaind^{d-1}#1\,\,}{\!\plaind^{d-1}#1\,\,}}
\newcommand{\ddMoint}[1]{\ddMOneint{#1}}

\newcommand{\dbar}{\plaind\mkern-6mu\mathchar'26}
\newcommand{\deltabar}{\delta\mkern-6mu\mathchar'26}
\newcommand{\dintbar}[1]{\mathchoice{\!\dbar#1\,}{\!\dbar#1\,}{\!\dbar#1\,}{\!\dbar#1\,}}
\newcommand{\ddintbar}[1]{\mathchoice{\!\dbar^d#1\,}{\!\dbar^d#1\,}{\!\dbar^d#1\,}{\!\dbar^d#1\,}}

\newcommand{\dTwointbar}[1]{\mathchoice{\!\dbar^{2}#1\,}{\!\dbar^{2}#1\,}{\!\dbar^{2}#1\,}{\!\dbar^{2}#1\,}}
\newcommand{\Dint}[1]{\mathcal{D}\!#1\,}

\usepackage{dsfont}

\newcommand{\canetset}[1]{{\mathchoice {\hbox{$\sf\textstyle #1\kern-0.4em #1$}}
{\hbox{$\sf\textstyle #1\kern-0.4em #1$}}
{\hbox{$\sf\scriptstyle #1\kern-0.3em #1$}}
{\hbox{$\sf\scriptscriptstyle #1\kern-0.2em #1$}}}}

\def\nbZ{{\mathchoice {\hbox{$\sf\textstyle Z\kern-0.4em Z$}}
{\hbox{$\sf\textstyle Z\kern-0.4em Z$}}
{\hbox{$\sf\scriptstyle Z\kern-0.3em Z$}}
{\hbox{$\sf\scriptscriptstyle Z\kern-0.2em Z$}}}}

\newcommand{\gpvec}[1]{\mathbf{#1}}

\newcommand{\zerovec}{\gpvec{0}}
\newcommand{\nullvec}{\zerovec}

\newcommand{\dvec}{\gpvec{d}}

\newcommand{\kvec}{\gpvec{k}}

\newcommand{\vvec}{\gpvec{v}}

\newcommand{\xvec}{\gpvec{x}}
\newcommand{\yvec}{\gpvec{y}}
\newcommand{\zvec}{\gpvec{z}}

\newcommand{\ident}{\mathds{1}}

\renewcommand{\AC}{\mathcal{A}}

\newcommand{\GC}{\mathcal{G}}

\newcommand{\OC}{\mathcal{O}}

\newcommand{\phitilde}{\tilde{\phi}}

\newcommand{\half}{\mathchoice{\frac{1}{2}}{(1/2)}{\frac{1}{2}}{(1/2)}}

\newcommand{\Exp}[1]{\operatorname{exp}\left(#1\right)}
\renewcommand{\exp}[1]{\mathchoice{\mathrm{e}^{#1}}{\operatorname{exp}\left(#1\right)}{\operatorname{exp}\left(#1\right)}{\operatorname{exp}\left(#1\right)}}

\renewcommand{\Dot}[1]{\mathring{#1}}

\newcommand{\elabel}[1]{\label{eq:#1}}
\newcommand{\eref}[1]{(\ref{eq:#1})}
\newcommand{\Eref}[1]{Eq.~(\ref{eq:#1})}
\newcommand{\Erefs}[1]{Eqs.~(\ref{eq:#1})}

\newcommand{\sref}[1]{Sec.~\ref{sec:#1}}
\newcommand{\Sref}[1]{Section~\ref{sec:#1}}

\newcommand{\Fref}[1]{Figure~\ref{fig:#1}}

\newcommand{\latin}[1]{{\it #1}}
\newcommand{\ie}{\latin{i.e.}\@\xspace}

\newlength \standardfigwidth
\DeclareMathAlphabet{\matheub}{U}{eur}{m}{n}

\newcounter{exercise}
{\addtocounter{exercise}{1}\begin{center}\begin{minipage}{0.8\linewidth}\textbf{Exercise
\arabic{exercise}:}\begin{itshape}}
{\end{itshape}\end{minipage}\end{center}}

\makeatletter
\newcommand{\creat}[3][]{\@ifempty{#1}{#2^{\dagger}}{\left(#2^{\dagger}\right)^{#1}}\@ifempty{#3}{}{\!(#3)}}

\newcommand{\creatDoi}[3][]{\@ifempty{#1}{\tilde{#2}}{\left(\tilde{#2}\right)^{#1}}\@ifempty{#3}{}{(#3)}}

\newcommand{\annih}[3][]{#2\@ifempty{#1}{}{^{#1}}\@ifempty{#3}{}{(#3)}}

\makeatother

% ...

\newlength{\bibmarkkeyAleft}

\newlength{\bibmarkkeyBleft}

\newlength{\bibmarkkeyCleft}

\newlength{\bibmarkkeyDleft}

\newcommand{\corresponds}{\mathrel{\hat{=}}}
\newcommand{\rightmoveprop}{\rightmovepropX{}{}}

\newcommand{\rightmovepropX}[2]{\tikz[baseline=-2.5pt]{
\draw[Aactivity] (180:0.5) -- (0,0)   node[at start, above] {$#1$};
\draw[Aactivity] (0:0.5) -- (0,0) node[at start,below] {$#2$};}}

\newcommand{\blobAmputated}[2]{
\tikz[baseline=-2.5pt]{
\draw[Aactivity]  (0,0)--(180:0.3)  node[at end ,above]{$#1$};
\draw[Aactivity]  (0,0)--(0:0.3)  node[at end ,above]{$#2$};
\draw[red,fill=red] (0,0) circle (1mm);
}}

\date{}
\title{Field Theory of Free Run and Tumble Particles in $d$ Dimensions }
\begin{document}
\author[a]{Ziluo Zhang}
\author[a]{Gunnar Pruessner}
\affil[a]{Department of Mathematics, Imperial College London, 180 Queen's Gate, London SW7 2AZ}
\renewcommand*{\Affilfont}{\small\it}
\maketitle

\begin{abstract}
In this paper, Doi-Peliti field theory is used to describe the motion of free Run and Tumble particles in arbitrary dimensions. After deriving action and propagators,
the mean square displacement and the corresponding entropy production at stationarity are calculated in this framework. We further derive the field theory of free Active Brownian Particles in two dimensions for comparison.
\end{abstract}
\section{Introduction}

The Run and Tumble (RnT) process is a random walk used to describe the motion of several biological species, such as \textit{Escherichia coli}, also known as \textit{E.~coli} \cite{deGennes2004}. Due to the lack of sensing organs, an \textit{E.~coli} performs a run with nearly constant speed $v$ in an almost straight line to probe the environment, until, with the Poissonian rate $\tumbleRate$, it parks for some time to choose a random direction from a uniform distribution to continue a ballistic motion \cite{SMRNT}. 
Any particle that follows an RnT process is called an RnT particle, also  known as persistent Brownian motion \cite{RNT1D}.

Doi-Peliti field theory \cite{Doi:1976, Peliti:1985} is a class of field theories  to deal with non-equilibrium statistical mechanics \cite{JPTHESIS}. Observables in this field theory are calculated from the master or Fokker-Planck equation \cite{TaeuberHowardVollmayr-Lee:2005,Garcia-MillanPruessner:2021}, applying second quantisation and the path integral, in a straight forward manner. The aim in the following is to apply Doi-Peliti field theory to the one and higher dimensional RnT process, to calculate the particle density subject to initial conditions, the propagators and to use it to determine the mean squared displacement (MSD) as well as the entropy production. A free RnT particle can take two directions in one dimension, so that there is always one direction available to be picked, which is inverse to the current direction. In one dimension, particles can be assigned two distinct species, depending on their drift direction. However, because of the infinite choices of directions in two  and higher dimensions, this option is no longer available as the range of species is continuous and periodic.

There are several questions we are motivated by. Firstly, we want to establish the field-theoretic framework to capture RnT motion in arbitrary dimensions exactly. While this is a mere exercise in one dimension, the two-dimensional problem is significantly richer. At this stage, we consider only the free case, \ie RnT particle without external potential. Secondly, we want to compare RnT motion to diffusive processes. While it is clear that the MSD is asymptotically identical to that of diffusion, the time and length scale below which this precisely breaks down is relevant in theories and applications. 
Thirdly, we want to determine the entropy production of free RnT particles in dimensions beyond one to see how previous results generalise.

In the following, we will derive the action and the propagator of one, two and higher dimensional RnT processes in \Sref{field-theory}, determine the MSD and the entropy production in \Sref{observables}, and the MSD of active Brownian particles in \Sref{ABP}. We discuss the results in \Sref{discussion}.

\section{Field theory of RnT particles}
\label{sec:field-theory}
The motion of RnT particles can be divided into two parts: ‘run’ and ‘tumble’. During a run, particles drift 
in the direction of a director
with constant speed $v$ deviating from a straight line only due to diffusion with constant $D$, until, during a 'tumble', they change direction instantaneously and spontaneously as the effect of an underlying Poisson process with rate $\tumbleRate$. In other words, particles tumble with rate $\tumbleRate$ and at tumbling, they immediately choose a new, random, uniformly distributed direction. In one dimension the director can take only two values, which is conveniently encoded by different particle species.
In two dimensions the director is parameterised by the angle $\theta$ relative to the $x$-axis and is, at tumbling, drawn uniformly from the interval $[0,2\pi)$. \Fref{rnt} shows a sample path of a two-dimensional RnT particle.

\begin{figure}[h]
\centering
\includegraphics[width=12cm,
  height=6cm,
  keepaspectratio]{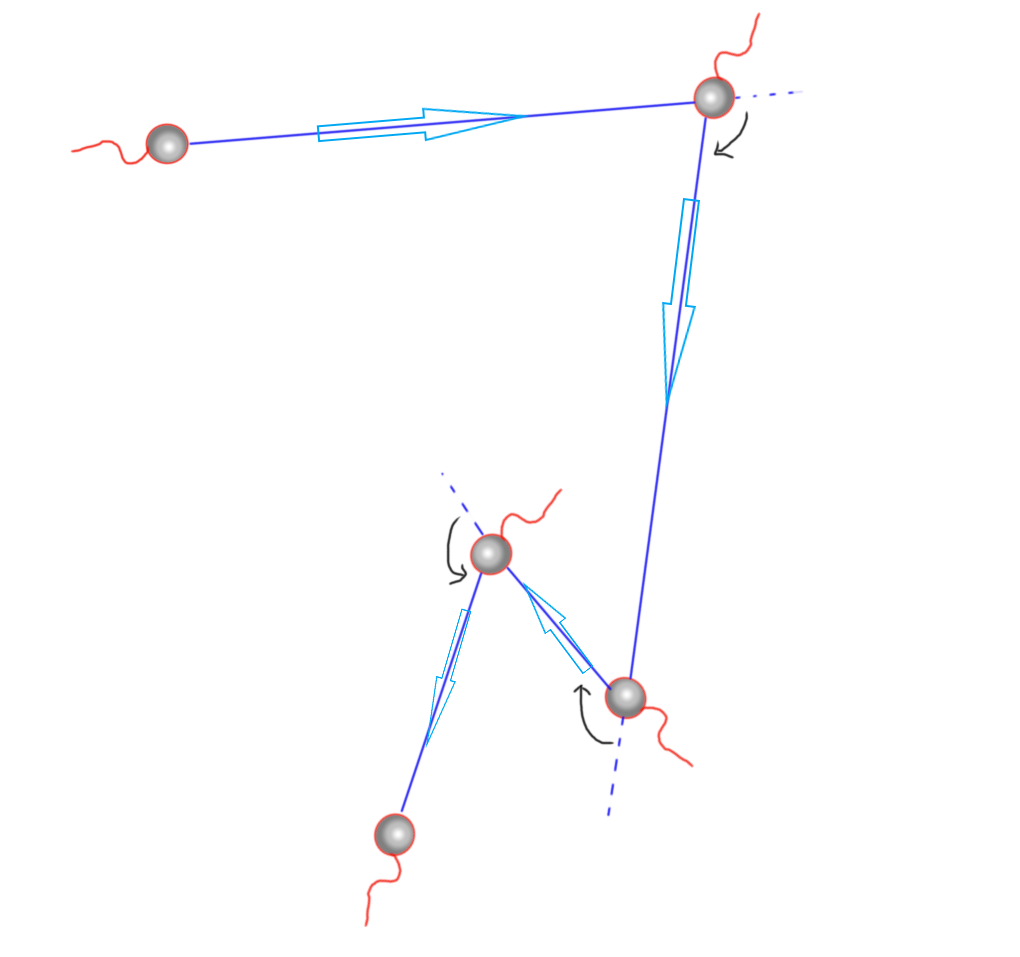}
\caption{An example of a 2D RnT particle's motion. The straight lines show the trajectory of the motion, the wriggly line is the 'tail' of the particle, which points the opposite direction of the velocity. The arrow from the dashed line to the straight line shows a change of the velocity during a tumble.}
\label{fig:rnt}
\end{figure}

In the following, we will derive the field theory of RnT particles. To illustrate the procedure, we will initially focus on the one-dimensional case. We will then derive the action in two dimensions and derive the propagator. Throughout we will use the following conventions for the Fourier-transform, of any annihilator field $\phi(\xvec,\theta,t)$ at position $\xvec$, director $\theta$ and time $t$
\begin{equation}
    \phi(\kvec,\theta,\omega)=\int \ddint{x} \dint{t}
    \exp{\imag\omega t}\exp{-\imag\kvec\xvec}
    \phi(\xvec,\theta,t) 
    \qquad \text{and} \qquad
    \phi(\xvec,\theta,t)=\int\ddintbar{k} \dintbar{\omega}
    \exp{-\imag\omega t}\exp{\imag\kvec\xvec}
    \phi(\kvec,\theta,\omega)\ ,
\end{equation}
where $\dbar=\plaind/(2\pi)$ and correspondingly for powers of the measure. Similarly, we use a matching $\deltabar(\omega)=2\pi \delta(\omega)$ and
$\deltabar(\kvec)=(2\pi)^d \delta(\kvec)$, so that powers of $2\pi$ never feature as a result of the Fourier transform. The Fourier transform of the creator fields is identical to that of the annihilator field.

Much of the focus of this work is on the full propagator
\begin{equation}\elabel{def_GC}
    \GC(\xvec-\xvec_0,\theta,\theta_0, t-t_0)=
    \ave{\phi(\xvec,\theta,t)\phitilde(\xvec_0,\theta_0,t_0)}
\end{equation}that is the probability of finding a particle at position $\xvec$ with director $\theta$ at time $t$ given one has been placed at $\xvec_0$ with director $\theta_0$ at time $t_0$. In the field theory, this probability is strictly speaking a particle density, given that only one particle has been created in a vacuum state.

The MSD can then be calculated by 
\begin{equation}
    \spave{\xvec^2}(t) = \SumInt_{\theta}
    \int \ddint{x} |\xvec|^2
    \GC(\xvec,\theta, \theta_0,t)
\end{equation}
choosing a convenient starting position, director and time. The sum or integral over $\theta$ is necessary to collect all contributions to the MSD irrespective of the director. After Fourier-transforming, the MSD can be extracted more efficiently, via 
\begin{equation}\elabel{MSD_from_GC}
    \spave{\xvec^2}(t) = 
    -\nabla^2_{\vec{k}=0}
    \SumInt_{\theta}
    \int \dintbar{\omega} \exp{-\imag\omega t}
    \GC(\kvec,\theta,\theta_0,\omega).
\end{equation}

\subsection{One dimension}
In one dimension, particles have only two velocities to drift, leftwards $-v$ and rightwards $v$ and therefore particles are in the Markovian sense fully characterised by their position $x$ and direction of the drift. The tumbling may be interpreted as a transmutation of a species 
drifting with velocity $v$ to another one with velocity $-v$ and vice versa. Because tumbling results in a randomly chosen direction and transmutation is a reversal of the director, the latter takes place with a rate half that of the former. The coupled Fokker-Planck equations of this process read
\begin{subequations}
\elabel{FPEs_1D}
\begin{align}
\partial_t P_L(x,t)&=D\partial_x^2 P_L(x,t) +v\partial_x  P_L(x,t)-\frac{\tumbleRate}{2} P_L(x,t)+\frac{\tumbleRate}{2} P_R(x,t),\\
\partial_t P_R(x,t)&=D\partial_x^2 P_R(x,t) -v\partial_x  P_R(x,t)-\frac{\tumbleRate}{2} P_R(x,t)+\frac{\tumbleRate}{2} P_L(x,t),
\end{align}
\end{subequations}
where $P_L(x,t)$ and $P_R(x,t)$ are the probability densities of leftward and rightward particle at position $x$ and time $t$ respectively. To pave the way for the two-dimensional case, we may write \Eref{FPEs_1D} in the form
\begin{align}\elabel{FPE_1D_better}
    \partial_t P(x,\theta_i,t) 
    &=D\partial_x^2 P(x,\theta_i,t) - v_i\partial_x  P(x,\theta_i,t) 
    - \tumbleRate P(x,\theta_i,t)
    + \frac{\tumbleRate}{M} \sum_{j=1}^M P(x,\theta_j,t) 
\end{align}
for a density $P(x,\theta_i,t)$ of finding a particle at time $t$ at $x$ with director $\theta_i$
where $\theta_i$ with $i\in\{1,\ldots,M\}$ refers to the $M=2$ directions particles can be moving in with velocity $v_1=v$ and $v_2=-v$. This probability density is in fact the full propagator \Eref{def_GC},
\begin{equation}
    \GC(\xvec-\xvec_0,\theta_i,\theta_j, t-t_0)=
    \ave{\phi(\xvec,\theta_i,t)\phitilde(\xvec_0,\theta_j,t_0)}=P(\xvec,\theta_i,t)
\end{equation}
with the initial condition stated explicitly. The field theory provides a perturbative approach to solving the Fokker-Planck equation.

With drift, diffusion and transmutation well established 
\cite{TaeuberHowardVollmayr-Lee:2005,Cardy:2008}, the action follows immediately
\begin{equation}
    \action\big([\phi(x,\theta_i,t),\phitilde(x,\theta_i,t)]\big)
    = \action_0\big([\phi(x,\theta_i,t),\phitilde(x,\theta_i,t)]\big) + \actionPert\big([\phi(x,\theta_i,t),\phitilde(x,\theta_i,t)]\big)
\end{equation} 
with a bilinear part $\action_0$ whose path integral can be taken easily and a  perturbative part $\actionPert$,
\begin{subequations}
\begin{align}
    \action_0\big([\phi(x,\theta_i,t),\phitilde(x,\theta_i,t)]\big)&=\int \dint{x}\dint{t} \sum_{i=1}^M 
    \phitilde(x,\theta_i,t)\big[\partial_t+v_i\partial_x-D\partial_x^2+\tumbleRate + r\big]\phi(x,\theta_i,t)
    \\
    \actionPert\big([\phi(x,\theta_i,t),\phitilde(x,\theta_i,t)]\big)&= - \frac{\tumbleRate}{M}
    \int \dint{x}\dint{t} \sum_{i,j=1}^M
    \phitilde(x,\theta_j,t) \phi(x,\theta_i,t)
\elabel{def_Apert_1D}\ ,
\end{align}
\end{subequations}
where $\phi$ and $\phitilde$ are the annihilation
and Doi-shifted \cite{Cardy:2008} creation fields respectively. The perturbation theory remains valid even when $D=0$, \ie results are expected to be analytical for all $D\ge0$.
To maintain causality, we have added additional mass $r>0$ to the bilinear part that corresponds to spontaneous extinction. This mass will be taken to $0$ once no longer needed. 

Any expectation is now calculated perturbatively about the bilinear part,
\begin{equation}
    \ave{\bullet}
    = \int \Dint{\phitilde} \Dint{\phi} \bullet \exp{-\action[\phitilde,\phi]} 
    = \ave{\bullet \exp{-\actionPert[\phitilde,\phi]}}_0 
    \quad\text{ with }\quad
    \ave{\bullet}_0
    = \int \Dint{\phitilde} \Dint{\phi} \bullet \exp{-\action_0[\phitilde,\phi]} \ .
\end{equation}
To render the action local, \ie to remove the time and space derivative, we perform a (unitary) Fourier transform, resulting in 
\begin{subequations}
\begin{align}\elabel{def_action0_1D}
    \action_0
    &=\int \dintbar{k}\dintbar{\omega} \sum_{i=1}^M 
    \phitilde(-k,\theta_i,-\omega)\big[-\imag \omega  + \imag v_i k + D k^2 + \tumbleRate + r]\phi(k,\theta_i,\omega)
    \\
    \actionPert
    &= - \frac{\tumbleRate}{M}
    \int \dintbar{k}\dintbar{\omega} \sum_{i,j=1}^M
    \phitilde(-k,\theta_j,-\omega) \phi(k,\theta_i,\omega)
\elabel{def_actionPert}\ ,
\end{align}
\end{subequations}
so that the bare propagators can be read off immediately,
\begin{align}
    \ave{ \phi(k,\theta_i,\omega) \phitilde(k',\theta_j,\omega') }_0 &=
    \frac{\deltabar(k+k^\prime)\delta_{i,j}\deltabar(\omega+\omega^\prime)}
    {-\imag\omega+\imag v_i k+D k^2+\tumbleRate + r}
    =:
    G(k,\theta_i,\omega) \deltabar(k+k^\prime)\delta_{i,j}\deltabar(\omega+\omega^\prime)
    \\
    &\corresponds
    \rightmovepropX{k,i,\omega}{k',j,\omega'} \ ,
\end{align}
where we have introduced the bare propagator as a diagram, to be read from right to left. Below, we will use the diagonal matrix of bare propagators, $\Gmatrix(k,\omega):=\diag(G(k,\theta_1,\omega),\ldots,G(k,\theta_M,\omega))$.

\Eref{def_actionPert} produces the perturbative transmutation 
\begin{equation}\elabel{perturbative_vertex_1D}
    \frac{\tumbleRate}{M} 
    \deltabar(k+k^\prime)\deltabar(\omega+\omega^\prime)
    \corresponds  \blobAmputated{i}{j}
\end{equation}
corresponding to a matrix $\Tmatrix$ with each element equal to $(\Tmatrix)_{ij}=\tumbleRate/M$, so that \Eref{def_actionPert} can be written as $\actionPert = - \int \dint{x}\dint{t} \sum_{i,j=1}^M
    \phitilde(-k,\theta_j,-\omega) (\Tmatrix)_{ji} \phi(k,\theta_i,\omega)$.

Rather than deploying the perturbative expansion, 
it is obviously much more straight forward to write the entire action for $M=2$ in  the form
\begin{multline}\elabel{action_directly}
    \action = \int \dintbar{k} \dintbar{\omega}
    \left(
    \begin{array}{c}
    \phitilde(-k,\theta_1,-\omega)\\
    \phitilde(-k,\theta_2,-\omega)
    \end{array}
    \right)
    \Amatrix
    \left(
    \begin{array}{c}
    \phi(k,\theta_1,\omega)\\
    \phi(k,\theta_2,\omega)
    \end{array}
    \right) \\
    \text{with}\qquad
    \Amatrix=\left(
    \begin{array}{cc}
    -\imag \omega + \imag v_1 k + D k^2 + \frac{M-1}{M} \tumbleRate +r & -\frac{\tumbleRate}{M}\\ 
    -\frac{\tumbleRate}{M} & -\imag \omega + \imag v_2 k + D k^2 + \frac{M-1}{M} \tumbleRate +r 
    \end{array}
    \right)
    \ ,
\end{multline}
so that with $v_1=v$, $v_2=-v$ and $r\to0$
\begin{multline}\elabel{full_propagator_1D_direct}
    \begin{pmatrix}
    \ave{\phi(k,\theta_1,\omega)\phitilde(k^\prime,\theta_1,\omega^\prime)}& \ave{\phi(k,\theta_1,\omega)\phitilde(k^\prime,\theta_2,\omega^\prime)}  \\ \ave{\phi(k,\theta_2,\omega)\phitilde(k^\prime,\theta_1,\omega^\prime)}& \ave{\phi(k,\theta_2,\omega)\phitilde(k^\prime,\theta_2,\omega^\prime)}
    \end{pmatrix}
=\deltabar(\omega+\omega^\prime)\deltabar(k+k^\prime) A^{-1}\\
=\frac{\deltabar(\omega+\omega^\prime)\deltabar(k+k^\prime)}{(-\imag\omega+Dk^2+\tumbleRate/2)^2+(vk)^2-(\tumbleRate/2)^2}\begin{pmatrix}
-\imag\omega-\imag k v+Dk^2+\tumbleRate/2 &\tumbleRate/2\\
\tumbleRate/2&-\imag\omega+\imag k v+Dk^2+\tumbleRate/2 \end{pmatrix}.
\end{multline}
These full propagators are to be recovered in the following, using the perturbative expansion
\begin{multline}\elabel{Dyson_sum_1D}
    \begin{pmatrix}
    \ave{\phi(k,\theta_1,\omega)\phitilde(k^\prime,\theta_1,\omega^\prime)}& \ave{\phi(k,\theta_1,\omega)\phitilde(k^\prime,\theta_2,\omega^\prime)}  \\ \ave{\phi(k,\theta_2,\omega)\phitilde(k^\prime,\theta_1,\omega^\prime)}& \ave{\phi(k,\theta_2,\omega)\phitilde(k^\prime,\theta_2,\omega^\prime)}
    \end{pmatrix}
= 
\deltabar(k+k^\prime)\deltabar(\omega+\omega^\prime)
( \Gmatrix + \Gmatrix \Tmatrix \Gmatrix + \ldots )
\\
=
\deltabar(k+k^\prime)\deltabar(\omega+\omega^\prime)
\Gmatrix ( \ident - \Tmatrix \Gmatrix )^{-1}
=
\deltabar(k+k^\prime)\deltabar(\omega+\omega^\prime)
( \Gmatrix^{-1} - \Tmatrix )^{-1} \ ,
\end{multline}
reproducing \Eref{full_propagator_1D_direct} as
\begin{equation}
    \Gmatrix^{-1} - \Tmatrix = 
    \left(
    \begin{array}{cc}
    -\imag \omega + \imag v_1 k + D k^2 +  \tumbleRate +r - \frac{\tumbleRate}{M} & -\frac{\tumbleRate}{M}\\ 
    -\frac{\tumbleRate}{M} & -\imag \omega + \imag v_2 k + D k^2 + \tumbleRate +r - \frac{\tumbleRate}{M} 
    \end{array}
    \right)
    = \Amatrix \ ,
\end{equation}
identical to the matrix used in \Eref{action_directly}. 

The pattern we follow below
to calculate the propagator in higher dimensions, in particular \Erefs{def_M} and \eref{full_propagator_2D},
corresponds in one dimension to writing 
the matrix of propagators \Eref{Dyson_sum_1D} in the form 
$\Gmatrix + \Gmatrix\Tmatrix\Gmatrix + \ldots = \Gmatrix + \Gmatrix\Tmatrix\Gmatrix (\ident - \Tmatrix\Gmatrix)^{-1}$, which corresponds to 
separating propagation without tumbling from any propagation with tumbling. 
We therefore calculate explicitly 
\begin{equation}
    \Gmatrix\Tmatrix\Gmatrix = 
    \frac{\tumbleRate}{2} 
    \left(
    \begin{array}{cc}
    G(k,\theta_1,\omega)G(k,\theta_1,\omega)&G(k,\theta_1,\omega)G(k,\theta_2,\omega)\\
    G(k,\theta_2,\omega)G(k,\theta_1,\omega)&G(k,\theta_2,\omega)G(k,\theta_2,\omega)
    \end{array}
    \right) \ .
\end{equation}
and
\begin{equation}
    \Gmatrix\Tmatrix\Gmatrix (\ident - \Tmatrix\Gmatrix)^{-1}
    =
    \frac{1}{1-\frac{\tumbleRate}{2}(G(k,\theta_1,\omega)+G(k,\theta_2,\omega))}
    \Gmatrix\Tmatrix\Gmatrix  \ ,
\end{equation}
so that in fact the full propagators according to \Eref{Dyson_sum_1D} are
\begin{multline}\elabel{full_propagator_1D}
\GC(k,\theta_i,\theta_j,\omega)\deltabar(k+k')\deltabar(\omega+\omega^\prime):=
\ave{\phi(k,\theta_i,\omega)\phitilde(k',\theta_j,\omega')}\\
= G(k,\theta_i,\omega) \deltabar(k+k')\deltabar(\omega+\omega^\prime) \Big(
\delta_{i,j} +
\frac{\tumbleRate}{2} G(k,\theta_j,\omega) \frac{1}{1-\TC_1(k,\omega)}
\Big)
\ ,
\end{multline}
with 
\begin{equation}\elabel{def_M1}
    \TC_1(k,\omega) := \frac{\tumbleRate}{2}\sum_{\ell=1}^2 G(k,\theta_\ell,\omega) \ .
\end{equation}
This concludes the derivation of the full propagator of RnT particles in one dimension.

To calculate the MSD below, we will draw on the density of finding a particle irrespective of its director, obtained by summing a column in \Eref{full_propagator_1D_direct}, for example for initial orientation $\theta_1$ (left column),
\begin{multline}
\ave{\big( \phi(k,\theta_1,\omega)
+ \phi(k,\theta_2,\omega) \big)
\phitilde(k^\prime,\theta_1,\omega^\prime)} \\= 
\deltabar(\omega+\omega^\prime)\deltabar(k+k^\prime)
\frac{-\imag \omega - \imag v k + Dk^2 + \tumbleRate}{(-\imag\omega+Dk^2+\tumbleRate/2)^2+(vk)^2-(\tumbleRate/2)^2}
\end{multline}
or more generally on the basis of \Eref{full_propagator_1D}
\begin{equation}\elabel{summed_full_propagator_1D}
    \sum_i \ave{\phi(k,\theta_i,\omega)\phitilde(k',\theta_j,\omega')} =
    \deltabar(k+k')\deltabar(\omega+\omega^\prime)
    \frac{G(k,\theta_j,\omega)}{1-\TC_1(k,\omega)}.
\end{equation}

\subsection{Two dimensions}
In higher dimensions, $d=2$ in the following, before generalising,
the Fokker-Planck equation \Eref{FPE_1D_better} is naturally extended to
\begin{align}\elabel{FPE_2D}
    \partial_t P(\xvec,\theta,t) 
    &=D
    \nabla_{\xvec}^2 P(\xvec,\theta,t) 
    - \vvec_{\theta}\cdot\nabla_{\xvec}  P(\xvec,\theta,t) 
    - \tumbleRate P(\xvec,\theta,t)
    + \frac{\tumbleRate}{2\pi} \int_0^{2\pi}\dint{\theta'} P(\xvec,\theta',t) \ .
\end{align}
The derivation of the propagator in two dimensions follows mostly the procedure in one dimension, but with 
$v_i\partial_x$ in \Eref{def_action0_1D} replaced by $\vvec_\theta\cdot\nabla_{\xvec}$ and 
the perturbative part of the action \Eref{def_Apert_1D} now reading
\begin{equation}
    \actionPert\big([\phi(\xvec,\theta,t),\phitilde(\xvec,\theta,t)]\big)= - \frac{\tumbleRate}{2\pi}
    \int \ddint{x}\dint{t} \int \dint{\theta} \dint{\theta'}
    \phitilde(\xvec,\theta,t) \phi(\xvec,\theta',t)
\elabel{def_Apert_2D}\ .
\end{equation}
The velocity $\vvec_\theta$ is now essentially given by a director $\dvec_\theta$,
\begin{equation}
    \vvec_\theta= v \dvec_\theta = v 
    \left(\!\!\begin{array}{c}
    \cos \theta\\
    \sin \theta
    \end{array}\!\!\right)
\end{equation}
and is thus a continuous function of $\theta\in[0,2\pi)$. The coupling between $\theta$ and $\kvec$ are therefore no longer a triviality. Still, the
bare propagator reads
\begin{align}\elabel{bare_propagator_2D}
    \ave{\phi(\kvec,\theta,\omega)\phitilde(\kvec',\theta',\omega')}_0
&= \frac{\deltabar^d(\kvec+\kvec^\prime)\delta(\theta-\theta')\deltabar(\omega+\omega^\prime)}
    {-\imag\omega+\imag \vvec_\theta \cdot \kvec +D \kvec^2+\tumbleRate + r}
    =
    G(\kvec,\theta,\omega) \deltabar^d(\kvec+\kvec^\prime)\delta(\theta-\theta')\deltabar(\omega+\omega^\prime)
    \\\nonumber
    &\corresponds
    \rightmovepropX{k,\theta,\omega}{k',\theta',\omega'} \ ,
\end{align}
with $\vvec_\theta\cdot\kvec$ coupling with $\theta$ and $\kvec$, while the 
the perturbative vertex is simply
\begin{equation}\elabel{perturbative_vertex_2D}
    \frac{\tumbleRate}{2\pi} 
    \deltabar^d(\kvec+\kvec^\prime)\deltabar(\omega+\omega^\prime)
    \corresponds  \blobAmputated{\theta}{\theta'} \ .
\end{equation}
As a result, the Dyson sum performed in \Eref{Dyson_sum_1D} no longer is a matter of simple matrix multiplication. Instead, the terms in
\begin{equation}
\ave{\phi(k,\theta,\omega)\phitilde(k',\theta',\omega')}
\corresponds
 \rightmoveprop
+ \rightmoveprop\!\!\!\!\blobAmputated{}{}\!\!\!\!\rightmoveprop 
+ \rightmoveprop\!\!\!\!\blobAmputated{}{}\!\!\!\!\rightmoveprop\!\!\!\!\blobAmputated{}{}\!\!\!\!\rightmoveprop 
+ \ldots
\end{equation}
have to be inspected more closely. Writing them out one by one, they are
\begin{subequations}
\elabel{2D_propagators}
\begin{align}
    \rightmoveprop &\corresponds G(\kvec,\theta,\omega) \deltabar^d(\kvec+\kvec^\prime)\delta(\theta-\theta')\deltabar(\omega+\omega^\prime) \elabel{explicit_straight} \\
    \rightmoveprop\!\!\!\!\blobAmputated{}{}\!\!\!\!\rightmoveprop & \corresponds G(\kvec,\theta,\omega)  \frac{\tumbleRate}{2\pi} G(\kvec,\theta',\omega) \deltabar^d(\kvec+\kvec^\prime)\deltabar(\omega+\omega^\prime) \elabel{explicit_one_tumble}\\
    \rightmoveprop\!\!\!\!\blobAmputated{}{}\!\!\!\!\rightmoveprop\!\!\!\!\blobAmputated{}{}\!\!\!\!\rightmoveprop & \corresponds G(\kvec,\theta,\omega)  \frac{\tumbleRate}{2\pi} \int \dint{\theta_1} G(\kvec,\theta_1,\omega) \frac{\tumbleRate}{2\pi}
    G(\kvec,\theta',\omega)
    \deltabar^d(\kvec+\kvec^\prime)\deltabar(\omega+\omega^\prime)\elabel{explicit_two_tumbles}
\end{align}
\end{subequations}
where the uniform density $1/(2\pi)$ plays, dimensionally, the same role as $\delta(\theta-\theta')$. These terms describe the following processes: \Eref{explicit_straight} is drift-diffusion without tumbling, so that the initial and final angles, $\theta'$ and $\theta$ respectively, are forced to equal. \Eref{explicit_one_tumble} describes a single tumble, allowing initial and final angles to take any value. The density of such events, as far as $\theta$ is concerned, is uniform $\frac{1}{2\pi}$. \Eref{explicit_two_tumbles} involves two tumbles, whereby the particle's director can point in any direction $\theta_1$ during the "intermezzo". 

While therefore the initial and final parts of the propagation have a pre-described direction, namely $\theta'$ and $\theta$, the direction of the intermediate stretches are independent. Although the sudden change of direction, which features in the field theory like a spontaneous teleportation from one $\theta$ to another, not dissimilar to resetting, initially poses a great inconvenience, in the present context it improves greatly the computability. Active Brownian particles, whose director is diffusive (\sref{ABP}), or RnT particles whose director is otherwise correlated, require a different treatment \cite{BothePruessner:2021}. 

The key-integral to perform is
\begin{equation}\elabel{def_M}
    \blobAmputated{}{}\!\!\!\!\rightmoveprop \corresponds
    \frac{\tumbleRate}{2\pi}\int_0^{2\pi} \dint{\theta_1} G(\kvec,\theta_1,\omega) = \TC_2(\kvec,\omega) \ ,
\end{equation}
so that the full propagator can in fact be written as
\begin{multline}\elabel{full_2D_propagator_in_M}
\ave{\phi(k,\theta,\omega)\phitilde(k',\theta',\omega')}
= G(\kvec,\theta,\omega) \deltabar^d(\kvec+\kvec^\prime)\delta(\theta-\theta')\deltabar(\omega+\omega^\prime) \\
+
G(\kvec,\theta,\omega)  \frac{\tumbleRate}{2\pi} G(\kvec,\theta',\omega) \deltabar^d(\kvec+\kvec^\prime)\deltabar(\omega+\omega^\prime)
\{1+\TC_2(\kvec,\omega)+\TC_2(\kvec,\omega)^2+\ldots\}\ .
\end{multline}
With a suitable choice of the coordinate system $\vvec_\theta \cdot \kvec = v |\kvec| \cos \theta$ as $k_y=0$ and with $z=\Exp{\imag\theta}$ therefore
\begin{multline}\elabel{TC2_final}
\TC_2(\kvec,\omega)= \frac{-\imag \tumbleRate}{2\pi} 
\ointctrclockwise_{|z|=1} \dint{z}
\frac{1}{(-\imag \omega + D\kvec^2+\tumbleRate+r)z+\half \imag v |\kvec| (z^2 + 1)}\\
=\frac{\tumbleRate}{\sqrt{(-\imag\omega+D\kvec^2+\tumbleRate+r)^2+\kvec^2 v^2}}
\end{multline}
where $\kvec^2=k_x^2+k_y^2$ and the branch of the square root runs along the negative real axis.
The geometric sum in \Eref{full_2D_propagator_in_M} produces $1/(1-\TC_2)$ as $|\TC_2|< 1$ for $\alpha>0$, so that finally the full propagator is
\begin{multline}\elabel{full_propagator_2D}
\GC(\kvec,\theta,\theta',\omega)\deltabar^d(\kvec+\kvec^\prime)\deltabar(\omega+\omega^\prime):=
\ave{\phi(\kvec,\theta,\omega)\phitilde(\kvec',\theta',\omega')}\\
= G(\kvec,\theta,\omega) \deltabar^d(\kvec+\kvec^\prime)\deltabar(\omega+\omega^\prime) \Big(
\delta(\theta-\theta') +
\frac{\tumbleRate}{2\pi} G(\kvec,\theta',\omega) \frac{1}{1-\TC_2(\kvec,\omega)}
\Big)
\ ,
\end{multline}
following the same structure as \Eref{full_propagator_1D}.
The first term contributing only for $\theta=\theta'$ is due to processes without a single tumble, which carry a "mass" of $\tumbleRate$.
\Eref{full_propagator_2D} is the full propagator of free RnT particles in two dimensions.

Again, for the MSD it will be more relevant to consider
\begin{equation}\elabel{summed_full_propagator_2D}
    \int \dint{\theta}
    \ave{\phi(k,\theta,\omega)\phitilde(k',\theta',\omega')}
=
\deltabar^d(\kvec+\kvec^\prime)\deltabar(\omega+\omega^\prime)
\frac{G(\kvec,\theta',\omega) }{1-\TC_2(\kvec,\omega)}
\end{equation}
by virtue of \Eref{def_M}. Again, \Eref{summed_full_propagator_2D} follows the same pattern as the result in one dimension, \Eref{summed_full_propagator_1D}.

\subsection{Higher dimensions}
Higher dimensions follow the pattern established above. In dimensions $d>2$ it is more convenient, however, to consider the annihilator field $\phi(\xvec,\vvec,t)$ a density in $d$-dimensional space $\xvec$ and $d-1$-dimensional velocity $\vvec$, as the velocity is constrained to the surface of a $d$-dimensional sphere. The harmonic part of the action then reads
\begin{equation}
    \action_0\big([\phi(x,\vvec,t),\phitilde(x,\vvec,t)]\big)=\int \ddint{x}\dint{t} \int_{|\vvec|=v} \ddMoint{v}
    \phitilde(\xvec,\vvec,t)\big[\partial_t+\vvec\cdot\nabla_{\xvec}-D\nabla_{\xvec}^2+\tumbleRate + r\big]\phi(\xvec,\vvec,t)
\end{equation}
and the perturbative action is
\begin{equation}
    \actionPert\big([\phi(\xvec,\vvec,t),\phitilde(\xvec,\vvec,t)]\big)= - \frac{\tumbleRate}{v^{d-1} S_d}
    \int \ddint{x}\dint{t} \int_{|\vvec|=v} \ddMoint{v} \int_{|\vvec'|=v} \ddMoint{v'}
    \phitilde(\vvec,\theta,t) \phi(\xvec,\vvec',t)
\elabel{def_Apert_dD}\ ,
\end{equation}
with the integrals in the velocity running over the a sphere that has an area of $v^{d-1}S_d$ with $S_d=2\pi^{d/2}/\Gamma(d/2)$, with each surface element having an area of $\ddMoint{v}$.

The higher-dimensional bare propagators are essentially identical to \Eref{bare_propagator_2D}
\begin{align}\elabel{bare_propagator_dD}
    \ave{\phi(\kvec,\vvec,\omega)\phitilde(\kvec',\vvec',\omega')}_0
&= \frac{\deltabar^d(\kvec+\kvec^\prime)\delta^{d-1}(\vvec-\vvec')\deltabar(\omega+\omega^\prime)}
    {-\imag\omega+\imag \vvec \cdot \kvec +D \kvec^2+\tumbleRate + r}\\
    &=
    G_d(\kvec,\vvec,\omega) \deltabar^d(\kvec+\kvec^\prime)\delta^{d-1}(\vvec-\vvec')\deltabar(\omega+\omega^\prime)
    \corresponds
    \rightmovepropX{k,\vvec,\omega}{k',\vvec',\omega'} \ ,
\end{align}
with the $\delta$-function $\delta^{d-1}(\vvec-\vvec')$ being a density function on the surface of a $d$-dimensional sphere, \ie it is a $\delta$-function in dimensional polar coordinates without the radial part, as $|\vvec|=|\vvec'|$. 

As done for \Eref{full_propagator_1D} and \eref{full_propagator_2D}, 
calculating the full propagator is a matter of integrating over the interaction vertex, which in dimensions $d\ge2$ gives \cite[App.~3]{LeBellac:1991}
\begin{multline}\elabel{generalised_TC}
    \blobAmputated{}{}\!\!\!\!\rightmoveprop 
    \corresponds
    \frac{\tumbleRate}{v^{d-1}S_d}\int_{|\vvec|=v}  \ddMoint{v} 
    G_d(\kvec,\vvec,\omega)
    =
    \frac{\tumbleRate}{S_d}\int_0^\pi  \dint{\theta} 
    \frac{S_{d-1} \sin^{d-2}\theta}{-\imag\omega + D\kvec^2+ \alpha + r + \imag v |\kvec| \cos \theta}\\
    =
    \frac{\tumbleRate\Gamma(d/2)}{v|\kvec|\sqrt{\pi}\Gamma((d-1)/2)}
    \int_0^{\pi} \dint{\theta} 
    \frac{\sin^{d-2}\theta}{\frac{-\imag\omega + D\kvec^2+ \alpha + r}{v|\kvec|}+\imag \cos\theta}
    =: \TC_d(\kvec,\omega) \ .
\end{multline}
This expression reproduces \Eref{def_M}, as $S_2=2\pi$ and $S_1=2$.
With \Eref{generalised_TC} in place, the full propagators in higher dimensions are
\begin{multline}\elabel{full_propagator_dD}
\GC_d(\kvec,\vvec,\vvec',\omega)\deltabar^d(\kvec+\kvec^\prime)\deltabar(\omega+\omega^\prime)=
\ave{\phi(\kvec,\vvec,\omega)\phitilde(\kvec',\vvec',\omega')}\\
= G_d(\kvec,\vvec,\omega) \deltabar^d(\kvec+\kvec^\prime)\deltabar(\omega+\omega^\prime) \Big(
\delta^{d-1}(\vvec-\vvec') +
\frac{\tumbleRate}{v^{d-1}S_d} G_d(\kvec,\vvec',\omega) \frac{1}{1-\TC_d(\kvec,\omega)}
\Big)
\ .
\end{multline}
The relevant expression for the MSD is generally of the form \Eref{summed_full_propagator_2D},
\begin{equation}\elabel{summed_full_propagator_dD}
        \int \ddMoint{v}
    \ave{\phi(k,\vvec,\omega)\phitilde(k',\vvec',\omega')}
=
\deltabar^d(\kvec+\kvec^\prime)\deltabar(\omega+\omega^\prime)
\frac{G_d(\kvec,\vvec',\omega) }{1-\TC_d(\kvec,\omega)} \ ,
\end{equation}
even in one dimension, \Eref{summed_full_propagator_1D}.
Above, we have calculated $\TC_1$, \Eref{def_M1}, and $\TC_2$, \Eref{TC2_final} explicitly.

\section{Observables}\label{sec:observables}
In the following, we calculate first the mean squared displacement of RnT particles in one and two dimensions, followed by their entropy production in various settings.
\subsection{Mean squared displacement}
The MSD is most conveniently calculated on the basis of \Eref{MSD_from_GC} and the general expression for the propagator integrated over the director, \Eref{summed_full_propagator_dD}. Taking the Laplacian inside the integral requires the Laplacian of $\GC_d$, which seems rather difficult to generalise, given that a general expression for $\TC_d$ is not available. However, derivatives and integration commute for $\TC_d$, \Eref{generalised_TC}, at this stage restricting the calculation to $d\ge2$. As $\TC_d(\kvec,\omega)$ is even in $\kvec$, the MSD has the simple form

\begin{equation}
\spave{\xvec^2}(t) =- 
\int \dintbar{\omega}\exp{-\imag\omega t}
\left(
    \frac{\left.\nabla_\kvec^2\right|_{\kvec=\nullvec} G_d(\kvec,\vvec',\omega)}{1-\TC_d(0,\omega)}
    +
    \frac{G_d(0,\vvec',\omega)\left.\nabla_\kvec^2\right|_{\kvec=\nullvec} \TC_d(\kvec,\vvec')}{(1-\TC_d(0,\omega))^2}
\right)
\end{equation}
Both derivatives are easy to determine, as \Eref{bare_propagator_dD} gives
\begin{equation}
    \left.\nabla_\kvec^2\right|_{\kvec=\nullvec} G_d(\kvec,\vvec',\omega) =
    - \frac{2Dd}{(-\imag \omega + \tumbleRate + r)^2} 
    - \frac{2v^2}{(-\imag \omega + \tumbleRate + r)^3} 
\end{equation}
and by taking this very same derivative in the first integral of \Eref{generalised_TC},
\begin{equation}
    \left.\nabla_\kvec^2\right|_{\kvec=\nullvec} \TC_d(\kvec,\omega) =
    \alpha \left.\nabla_\kvec^2\right|_{\kvec=\nullvec} G_d(\kvec,\vvec',\omega)
    \ ,
\end{equation}
as the integrand becomes independent of $\vvec$ at $\kvec=\nullvec$. Similarly $\TC_d(0,\omega)=\tumbleRate G_d(0,\vvec,\omega) = \tumbleRate (-\imag \omega + \tumbleRate + r)^{-1}$. In particular 
\begin{equation}
    \frac{1}{1-\TC_d(0,\omega)} = 
    \frac{G_d^{-1}(0,\vvec,\omega)}{-\imag \omega + r}
\end{equation}
so that after some algebra for $d\ge2$
\begin{align}
\spave{\xvec^2}(t) &= \int\dintbar{\omega} \exp{-\imag \omega t} 
\left( \frac{1}{-\imag \omega +r} + \frac{\alpha}{(-\imag \omega +r)^2} \right)
\left(
\frac{2Dd}{-\imag \omega + \tumbleRate + r}
+
\frac{2v^2}{(-\imag \omega + \tumbleRate + r)^2}
\right)
\nonumber
\\
&=2Ddt+2\frac{v^2}{\tumbleRate^2} 
\left( \exp{-\tumbleRate t} - 1 + \tumbleRate t \right)
\elabel{MSD_dD}
\ .
\end{align}
By direct calculation from \Erefs{def_M1} and \eref{summed_full_propagator_1D} this expression turns out to hold also for $d=1$. The MSD is thus identical to that of a diffusive particle at times $t\gg1/\tumbleRate$ with effective diffusion constant $D+v^2/(d\tumbleRate)$ \cite{Cates_2013}, which is one of the key results of this work. As the plain diffusion with constant $D$ is modelled to equally take place in all spatial directions, the diffusion constant enters into the MSD linearly in the dimensionality of space $d$. The ballistic MSD $v^2t^2$ is indeed visible as $\exp{-\tumbleRate t} - 1 + \tumbleRate t = (\tumbleRate t)^2/2 + \OC(t^3)$ at short times, provided $Dd$ is small enough, as diffusion otherwise dominates at the short timescale. That $D$ can be chosen to vanish is consistent with the perturbation theory, which does not rely on non-vanishing $D$.

\subsection{Entropy production rate}\label{sec:entropy_production}
The internal entropy production rate for RnT particles in $d\ge2$ dimensions can be calculated via \cite{Gaspard:2007,Cocconi:2020,Garcia-MillanPruessner:ToBePublished}
\begin{multline}\elabel{def_entropy_production}
    \Dot{S}(t):=\lim_{\tau\rightarrow 0}
    \int \ddint{x} \int \ddint{y} 
    \int_{|\vvec|=v} \ddMoint{v} \int_{|\vvec'|=v} \ddMoint{v'}\\
    \times P(\xvec,\vvec,t) \Dot{W}(\xvec,\vvec \rightarrow \yvec,\vvec';\tau)
    \ln \left(
    \frac
    {P(\xvec,\vvec,t) W(\xvec,\vvec \rightarrow \yvec,\vvec';\tau)}
    {P(\yvec,\vvec',t) W(\yvec,\vvec' \rightarrow \xvec,\vvec;\tau)}
    \right) \ ,
\end{multline}
where $P(\xvec,\vvec,t)$ denotes the probability density of finding the particle at position $\xvec$ and with velocity $\vvec$ at time $t$ and $W(\xvec,\vvec \rightarrow \yvec,\vvec';\tau)$ the probability density for a particle to be found at $\yvec$ with velocity $\vvec'$ given it was at $\xvec$ with velocity $\vvec$ time $\tau$ earlier. The product $PW$ is thus the joint probability. The dot in $\Dot{W}$ denotes the derivative with respect to time, so that $\Dot{W}$ is in fact a transition rate.
At stationarity the density $P$ can be shown to drop from the logarithm, so that the internal entropy production rate equals the negative of the external entropy production rate. 

The system studied in the present work is infinite and no stationary state exists. However, the entropy production rate is concerned only with the short-time behaviour of the system, so that we can use the transition densities, \ie the propagators, derived above to determine the stationary internal entropy production rate of a finite, periodically closed system with constant $P(\xvec,\vvec,t)=P_0$, so that
\begin{equation}\elabel{entropy_production_stationary}
    \Dot{S}(t)=\lim_{\tau\rightarrow 0}
    \int \ddint{x} \int \ddint{y} 
    \int_{|\vvec|=v} \ddMoint{v} \int_{|\vvec'|=v} \ddMoint{v'}
    P_0 \Dot{W}(\xvec,\vvec \rightarrow \yvec,\vvec';\tau)
    \ln \left(
    \frac
    {W(\xvec,\vvec \rightarrow \yvec,\vvec';\tau)}
    {W(\yvec,\vvec' \rightarrow \xvec,\vvec;\tau)}
    \right) \ .
\end{equation}
In one dimension, the entropy production is known to be $\Dot{S}(t)=v^2/D$ \cite{Cocconi:2020}. In one dimension, the expression to be calculated differs slightly from \Eref{def_entropy_production}, because the velocity is not continuous, so that the integrals have to be replaced by suitable sums. In the following, we focus on $d=2$ dimensions, but the result trivially generalises to $d\ge2$.

As the transition rate $\lim_{\tau\to0}\Dot{W}(\ldots;\tau)=\Dot{W}(\ldots;0)$ is in fact the kernel of the Fokker-Planck equation \cite{Garcia-MillanPruessner:2021,Cocconi:2020,Garcia-MillanPruessner:ToBePublished}, re-written to match the form of a master equation,
\begin{multline}\elabel{FPE_kernel}
    \partial_t P(\xvec,\theta,t) 
    =
    \int\dTwoint{y}\int_0^{2\pi}\dint{\theta'}
    P(\yvec,\theta',t)
    \Dot{W}(\yvec,\theta' \rightarrow \xvec,\theta;0)
    \\
    =
    \int\dTwoint{y}\int_0^{2\pi}\dint{\theta'}
    P(\yvec,\theta',t)
    \Bigg\{
    \delta(\theta-\theta')
    \Big[
    D \nabla_{\yvec}^2  
    +\vvec_{\theta'}\cdot\nabla_{\yvec} 
    - \tumbleRate 
    \Big] \delta(\xvec-\yvec)
    + \frac{\tumbleRate}{2\pi} \delta(\xvec-\yvec) 
    \Bigg\}\ ,
\end{multline}
the transition rate is simply  \cite{Garcia-MillanPruessner:ToBePublished}
\begin{equation}\elabel{opTerm}
   \Dot{W}(\yvec,\theta' \rightarrow \xvec,\theta;0) = 
    \delta(\theta-\theta')
    \Big[
    D \nabla_{\yvec}^2  
    + \vvec_{\theta'}\cdot\nabla_{\yvec} 
    - \tumbleRate 
    \Big] \delta(\xvec-\yvec)
    + \frac{\tumbleRate}{2\pi} \delta(\xvec-\yvec) 
    \ .
\end{equation}
As for the transition probabilities in the logarithm, it needs to be determined to leading order in $\tau$, for the limit $\tau\to0$ to be taken afterwards. One can show that terms to $n$th order in the perturbation are of order $\tau^n$  \cite{Garcia-MillanPruessner:ToBePublished}, which means that 
\begin{equation}
    \lim_{\tau\to0}
    \ln \left(
    \frac
    {W(\xvec,\theta \rightarrow \yvec,\theta';\tau)}
    {W(\yvec,\theta' \rightarrow \xvec,\theta;\tau)}
    \right)
=
    \lim_{\tau\to0}
    \ln \left(
    \frac
{     \rightmovepropX{\yvec,\theta',\tau}{\xvec,\theta,0}
+ \rightmovepropX{\yvec,\theta',\tau}{}\!\!\!\!\blobAmputated{}{}\!\!\!\!\rightmovepropX{}{\xvec,\theta,0} 
}
{     \rightmovepropX{\xvec,\theta,\tau}{\yvec,\theta',0}
+ \rightmovepropX{\xvec,\theta,\tau}{}\!\!\!\!\blobAmputated{}{}\!\!\!\!\rightmovepropX{}{\yvec,\theta',0} 
}
    \right) \ .
\end{equation}
with the diagrams given in \Eref{2D_propagators}. The inverse Fourier transform of \Eref{explicit_straight} gives a simple Gaussian, restricting, however, the final velocity to the initial one,  $\delta(\theta-\theta')(4D\tau)^{-d/2}\exp{-(\yvec-\xvec-\vvec_{\theta}\tau)^2/(4D\tau)}$, while the second term, \Eref{explicit_one_tumble}, 
 requires a double convolution over an intermediate time $t'$ and an intermediate position $\zvec$,
\begin{align}
    \rightmovepropX{\yvec,\theta',\tau}{}\!\!\!\!\blobAmputated{}{}\!\!\!\!\rightmovepropX{}{\xvec,\theta,0}
    &=
    \int \dTwoint{z}\int_0^\tau \dint{t'}
    \frac{1}{(4D t')^{d/2}}
    \exp{-\frac{(\zvec-\xvec-\vvec_{\theta} t')^2}{(4D t')}}
    \frac{\alpha}{2\pi}
    \frac{1}{(4D (\tau-t'))^{d/2}}
    \exp{-\frac{(\yvec-\zvec-\vvec_{\theta'} (\tau-t'))^2}{(4D (\tau-t'))}}
    \nonumber\\&=
    \frac{\alpha}{2\pi}
    \int_0^\tau \dint{t'}
    \frac{1}{(4D \tau)^{d/2}}
    \exp{-\frac{[\yvec-\xvec-(\vvec_{\theta} t' + \vvec_{\theta'} (\tau-t'))]^2}{4D \tau}}
\end{align}
The last integral can be carried out in closed form, but it is more instructive to expand the square in the potential and reparameterising, $s=(t'-\tau/2)/\tau$,
\begin{multline}
    \int_0^\tau \dint{t'}
    \exp{-\frac{[\yvec-\xvec-(\vvec_{\theta} t' + \vvec_{\theta'} (\tau-t'))]^2}{4D \tau}}
=
    \exp{-\frac{(\yvec-\xvec)^2 - (\yvec-\xvec)\cdot(\vvec_{\theta}+\vvec_{\theta'})\tau}{4D \tau}}
    \tau \int_{-1/2}^{1/2} \dint{s}
    \exp{\frac{(\yvec-\xvec)\cdot
    (\vvec_{\theta} - \vvec_{\theta'})s}{2D}}
    \exp{-\frac{[\vvec_{\theta} (1/2 + s) + \vvec_{\theta'} (1/2-s)]^2}{4D}\tau} \ .
\end{multline}
Writing the second exponential under the integral as $1+\OC(\tau)$ 
and that in turn as $\exp{-\frac{\left[\frac{\vvec_{\theta} + \vvec_{\theta'} }{2}\tau \right]^2}{4D\tau }}(1+\OC(\tau))$ to complete the square with the exponential prefactor,
leaves us with a simple exponential to be integrated, so that 
\begin{multline}
    \rightmovepropX{\yvec,\theta',\tau}{}\!\!\!\!\blobAmputated{}{}\!\!\!\!\rightmovepropX{}{\xvec,\theta,0}
    =
    \frac{\alpha\tau}{2\pi}
    (4D \tau)^{-d/2}
    \Exp{-\frac{\left[ \yvec-\xvec - \frac{\vvec_{\theta}+\vvec_{\theta'}}{2})\tau\right]^2}{4D \tau}}
    \\
    \times
    \frac{2D}{(\yvec-\xvec)\cdot(\vvec_{\theta} - \vvec_{\theta'})}
\left(
    \exp{\frac{(\yvec-\xvec)\cdot(\vvec_{\theta} - \vvec_{\theta'})}{4D}}
-
    \exp{-\frac{(\yvec-\xvec)\cdot(\vvec_{\theta} - \vvec_{\theta'})}{4D}}
\right) 
    (1+\OC(\tau)) \ .
\end{multline}
In small $\yvec-\xvec$ effectively considered in the integral over $\xvec$ and $\yvec$ in \Eref{entropy_production_stationary}, the fraction multiplying the difference of exponentials in the second line converges to $1$, so that finally
\begin{equation}
    \rightmovepropX{\yvec,\theta',\tau}{}\!\!\!\!\blobAmputated{}{}\!\!\!\!\rightmovepropX{}{\xvec,\theta,0}
    =
    \frac{\alpha\tau}{2\pi}
    (4D \tau)^{-d/2}
    \Exp{-\frac{\left[ \yvec-\xvec - \frac{\vvec_{\theta}+\vvec_{\theta'}}{2})\tau\right]^2}{4D \tau}}
    (1+\OC(\tau,|\yvec-\xvec|)) \ .
\end{equation}
In summary 
\begin{multline}
W(\xvec,\theta \rightarrow \yvec,\theta';\tau) 
=
    \rightmovepropX{\yvec,\theta',\tau}{\xvec,\theta,0}
+ \rightmovepropX{\yvec,\theta',\tau}{}\!\!\!\!\blobAmputated{}{}\!\!\!\!\rightmovepropX{}{\xvec,\theta,0} + \OC(\tau^2)
\\=
\frac{\delta(\theta-\theta')}{(4 D \tau)^{d/2}}\exp{-\frac{(\yvec-\xvec-\vvec_{\theta}\tau)^2}{4D\tau}}
+
    \frac{\alpha\tau}{2\pi}
    \frac{1}{(4D \tau)^{d/2}}
    \exp{-\frac{\left[ \yvec-\xvec - \left(\frac{\vvec_{\theta}+\vvec_{\theta'}}{2}\right)\tau\right]^2}{4D \tau}}
    \Big(1+\OC(\tau,|\yvec-\xvec|)\Big) \ .
\end{multline}
Taking the ratio of this expression and the one obtained by exchanging final and initial positions and velocities, results in a fraction containing a Dirac $\delta$-function of the angles in the denominator, which needs to be rendered meaningful, for example by discretising the angles and replacing the Dirac $\delta$ by a suitably weighted Kronecker $\delta$, whose weight means that if $\theta=\theta'$ the bare propagator dominates over the perturbative term, but subsequently cancels  from the fraction. It follows that
\begin{equation}
    \lim_{\tau\to0}
    \ln \left(
    \frac
    {W(\xvec,\theta \rightarrow \yvec,\theta';\tau)}
    {W(\yvec,\theta' \rightarrow \xvec,\theta;\tau)}
    \right)
=
\begin{cases}
\frac{(\yvec-\xvec) \cdot \vvec_{\theta}}{D} \text{ for } \theta=\theta' \\
\frac{(\yvec-\xvec) \cdot \frac{\vvec_{\theta}+\vvec_{\theta'}}{2}}{D} \text{ otherwise, }
\end{cases}
\end{equation}
which can be summarised by
\begin{equation}\elabel{lnTerm}
    \lim_{\tau\to0}
    \ln \left(
    \frac
    {W(\xvec,\theta \rightarrow \yvec,\theta';\tau)}
    {W(\yvec,\theta' \rightarrow \xvec,\theta;\tau)}
    \right)
=
\frac{1}{D}
(\yvec-\xvec) \cdot \frac{\vvec_{\theta}+\vvec_{\theta'}}{2}
\end{equation}
This expression applies across all dimensions, including $d=1$, where, however, the case of tumbling, $\vvec_{\theta}\ne\vvec_{\theta'}$ implies that $\vvec_{\theta}+\vvec_{\theta'}=0$, which is why terms with $\vvec_{\theta}\ne\vvec_{\theta'}$ cannot enter into the entropy production of RnT particles in one dimension at stationarity \cite{Cocconi:2020}.

Using \Eref{opTerm} in the required form 
and \Eref{lnTerm} in \Eref{entropy_production_stationary} then gives
\begin{multline}\elabel{entropy_production_rate_final}
    \Dot{S}(t)=
    \int \dTwoint{x} \dTwoint{y} 
    \int_0^{2\pi} \dint{\theta} \dint{\theta'}
P_0\frac{1}{D}
(\yvec-\xvec) \cdot \frac{(\vvec_{\theta}+\vvec_{\theta'})}{2}\\
\times
\left\{
    \delta(\theta-\theta')
    \Big[
    D \nabla_{\xvec}^2  
    + \vvec_{\theta}\cdot\nabla_{\xvec} 
    - \tumbleRate 
    \Big] \delta(\xvec-\yvec)
    + \frac{\tumbleRate}{2\pi} \delta(\xvec-\yvec) 
\right\}\\
=
    \int \dTwoint{x} \dTwoint{y} 
    \int_0^{2\pi} \dint{\theta} \dint{\theta'}
\frac{P_0}{D}
\delta(\xvec-\yvec)
    \delta(\theta-\theta')
\left[ -\vvec_{\theta}\cdot\nabla_{\xvec}\right]
(\yvec-\xvec) \cdot \frac{\vvec_{\theta}+\vvec_{\theta'}}{2}
=\frac{v^2}{D}
\end{multline}
using integration by parts and the normalisation of $P_0$ such that
$\int \dTwoint{x} \dTwoint{y} \int_0^{2\pi} \dint{\theta} \dint{\theta'} 
\delta(\theta-\theta')
\delta(\xvec-\yvec)
P_0=1$. The entropy production is therefore the same in all dimensions and does not depend on the tumble rate. The entropy production of the steady state may also be calculated via probability current \cite{Seifert:2005}, $\nabla_{\xvec} J=-\partial_t P_0$,
\begin{equation}
    \Dot{S}(t)=\int\dTwoint{x}\int_0^{2\pi}\dint \theta\frac{J^2}{DP_0}
    =\int\dTwoint{x}\int_0^{2\pi}\dint \theta\frac{(-D\nabla_\xvec P_0+\vvec_\theta P_0)^2}{DP_0}=\frac{v^2}{D}
\end{equation}
which is consistent with \Eref{entropy_production_rate_final}.

\section{Active Brownian Particles}\label{sec:ABP}
The active Brownian particle (ABP) model \cite{Cates_2013,PhysRevLett.121.078001,Stenhammar:2014,2102.13007} describes the motion of a particle in $d\ge2$ dimensions that moves with constant speed $|\vvec|=v$ but whose director undergoes diffusion with constant $D_r$ on a $d$-dimensional sphere. The particle might further diffuse itself with constant $D$. What makes the ABP model different from RnT is the continuous evolution of the director. In one dimension this is no longer possible and the ABP model may be thought of degenerating into RnT.

In particular, an ABP in two dimensions moves with velocity $\vvec_\theta=v[\cos\theta,\sin\theta]^T$, parameterised by an angle $\theta$ that undergoes rotational diffusion with $D_r$. 
In the following, we will derive the action, the propagator and the MSD of ABP in two dimensions, via the field theory, demonstrating  the equivalence between RnT and ABP as far as the MSD is concerned, both on the large and the small time scale.

The Fokker-Planck equation of ABP reads \cite{PhysRevLett.121.078001}
\begin{equation}
    \partial_t P(\xvec,\theta,t)=\bigg[D\nabla^2-\vvec_\theta \nabla+D_r\partial^2_\theta\bigg]P(\xvec,\theta,t) \ ,
\end{equation}
so that the action is \cite{Garcia-MillanPruessner:ToBePublished}
\begin{align}
    &\AC\bigg([\phi(\kvec,\theta,\omega),\phitilde(\kvec,\theta,\omega)]\bigg)\\&=\int_0^{2\pi}\dint{\theta}
    \int\dintbar{\omega}\dTwointbar{k} \phitilde(-\kvec,\theta,-\omega)\bigg[-\imag\omega+D\kvec^2+\imag vk\cos\theta-D_r\partial^2_{\theta} +r \bigg]\phi(\kvec,\theta,\omega)\elabel{originABPaction}.
\end{align}
after Fourier-transforming in space and time and after adding a mass $r>0$ to preserve causality. Following the procedure above, we would treat the angular diffusion term $\partial_{\theta}^2$ perturbatively. However, as discussed in the following, the differential operator in $\theta$ is essentially the Mathieu equation with known eigenfunctions and eigenvalues $\lambda_{2j}$ \cite{ZIENER2012,AbramowitzStegun}, so that $\AC$ can be made local in $\omega$, $\kvec$ and $\lambda_n$ simultaneously \cite{BothePruessner:2021}. This results in a particularly elegant action, which can readily be used in perturbative calculations. Alternatively Fourier-summing, so that $\phi(\kvec,\theta,\omega)=(2\pi)^{-1}\sum_n \exp{\imag n \theta} \phi_n(\kvec,\omega)$ results in the interaction term 
$\imag vk \cos\theta$ transforming to
$\imag vk \half (\delta_{n,m+1}+\delta_{n,m-1})$, allowing for terms of the form $\phitilde_n(-\kvec,-\omega)\phi_{n\pm1}(\kvec,\omega)$, which generally have to be treated perturbatively in a comparatively messy way.

\newcommand{\repang}{\gamma}
To proceed with the derivation in terms of Mathieu-functions, we reparameterise $\repang=\theta/2$, 
\begin{align}
&\AC\bigg([\phi(\kvec,\repang,\omega),\phitilde(\kvec,\repang,\omega)]\bigg)\\
&=2\int_0^{\pi}\dint{\repang} \int\dintbar{\omega}\dTwointbar{k} \phitilde(-\kvec,\repang,-\omega)\bigg[-\imag\omega+D\kvec^2+\imag vk\cos(2\repang)-\frac{1}{4}D_r\partial^2_{\repang}+r\bigg]\phi(\kvec,\repang,\omega)\elabel{ABPaction}
\end{align}
and introduce 
\begin{subequations}
\elabel{fields_in_u}
\begin{align}
\phi(\kvec,\repang,\omega)&=\sum_n \phi_n(\kvec,\omega)u_n(\repang,q)\\
\phitilde(\kvec,\repang,\omega)&=\sum_n \phitilde_n(\kvec,\omega)\Tilde{u}_n(\repang,q)
\end{align}
\end{subequations}
where $u_n(\repang,q)$ 
depend on the dimensionless, purely imaginary parameter $q=2\frac{\imag vk}{D_r}$
and are eigenfunctions of the angle-dependent part of the operator in \Eref{ABPaction},
\begin{equation}\elabel{Mathieu-op}
    \bigg[\partial_\repang^2-2 q\cos(2\repang))\bigg]u_n(\repang,q)=
    - \lambda_n(q) u_n(\repang,q) \ .
\end{equation}
The solutions of this eigenvalue problem are the even 
\begin{equation}\elabel{ce_Fourier}
      u_{2\ell}(\repang,q)=\ce_{2\ell}(\repang,q)=\sum_{j=0} A_{2j}^{(2\ell)}(q)\cos(2j\repang)
\end{equation}
and odd
\begin{equation}\elabel{se_Fourier}
    u_{2\ell+1}(\repang,q)=\se_{2\ell+2}(\repang,q)\sum_{j=0} B_{2j+2}^{(2\ell+2)}(q) \sin(2j\repang)
\end{equation}
$\pi$-periodic Mathieu-functions \cite{2102.13007,ZIENER2012,AbramowitzStegun}, which can be written as Fourier series as outlined above and whose eigenvalues have been characterised \cite{ZIENER2012} as expansions in $q$
\begin{equation}
    \lambda_n(q) = \sum_{j=0}^\infty q^j \alpha_{n,j} \ ,
\end{equation}
where, in particular,
$\lambda_0(q)=-q^2/2+\OC(q^4)$
and
$\lambda_2(q)=4+5q^2/12+\OC(q^4)$ in small $q$.
The Fourier coefficients of the even Mathieu-functions, \Eref{ce_Fourier}, in particular are also known \cite{ZIENER2012},
\begin{align}
    A_0^{(0)}(q)&=\frac{1}{\sqrt{2}}\bigg[1-\frac{q^2}{16}\bigg]+\mathcal{O}(q^3),\\
    A_0^{(2)}(q)&=\frac{q}{4}+\mathcal{O}(q^2)\\
    A_0^{(2j)}(q)&=\frac{1}{(2j-1)!j}\bigg[\frac{q}{4}\bigg]^j+\mathcal{O}(q^{j+1}),\text{for } j>0 \ ,
\end{align}
in small $q$.
As the operator \Eref{Mathieu-op} is self-adjoint, with suitable normalisation $\tilde{u}_n(\gamma,q)=u_n(\gamma,q)/\pi$ \cite{ZIENER2012}, these functions create an orthonormal system,
\begin{align}
    2\int_0^\pi \dint\repang u_n(\repang,q)\tilde{u}_m(\repang,q)
    &=\int_0^{2\pi}\dint\theta u_n(\theta/2,q)\tilde{u}_m(\theta/2,q)=\delta_{n,m} \ .
\end{align}
Using \Eref{fields_in_u} in the action \Eref{ABPaction} then results in the diagonal action
\begin{align}
     \AC&=2\int_0^{\pi}\dint \repang\sum_{n,m}\int\dbar\omega\dbar\kvec \phitilde_n(-\kvec,-\omega)\Tilde{u}_n(\repang,q)\bigg[-\imag\omega+D\kvec^2+\frac{1}{4}D_r\lambda_m(q)+r\bigg]\phi_m(\kvec,\omega)u_m(\repang,q)\\
     &=\sum_{n}\int\dbar\omega\dTwointbar{k} \phitilde_n(-\kvec,-\omega)\bigg[-\imag\omega+D\kvec^2+\frac{1}{4}D_r\lambda_n(q)+r\bigg]\phi_n(\kvec,\omega)
     \elabel{action_abp_final}
     \ .
\end{align}

The bare propagator of this field theory,  \Eref{action_abp_final}, can be identified immediately, 
\begin{equation}\elabel{ABP_propagator}
    \ave{\phi_n(\kvec,\omega)\phitilde_m(\kvec',\omega')}
    =
    \frac{\delta_{n,m}\deltabar(\omega+\omega')\deltabar^2(\kvec+\kvec')}{-\imag \omega + D \kvec^2 + D_r \lambda_n(q)/4+r}
\end{equation}
To calculate the MSD irrespective of the final direction and considering a uniform initial direction, the propagator 
\begin{equation}
    \ave{\phi(\kvec,\theta,\omega)\phitilde(\kvec',\theta',\omega')}
    =
    \sum_{n,m}
    \ave{\phi_n(\kvec,\omega)\phitilde_m(\kvec',\omega')}
    u_n(\theta/2)\tilde{u}_m(\theta'/2)
\end{equation}
has to be integrated over all $\theta$ and averaged over $\theta'$, which is simplified greatly by observing that 
$\int_0^{2\pi}\dint{\theta} \sin(j\theta)=0$ for all $j\in \mathbb{Z}$, 
and
$\int_0^{2\pi}\dint{\theta} \cos(j\theta)=0$ for $\mathbb{Z}\ni j\ne0$, 
so that
integrals of $u_n$ for odd indeces $n$ \Eref{se_Fourier} vanish and those for even indeces produce $2\pi$ for each $0$-mode,
\begin{equation}
    \int_0^{2\pi} \dint{\theta} u_{2\ell} (\theta/2) 
    = 2\pi A_{0}^{(2\ell)}(q)
\end{equation}
It follows that
\begin{equation}
\frac{1}{2\pi}
    \int_0^{2\pi} \dint{\theta'} 
    \int_0^{2\pi} \dint{\theta}
    \ave{\phi(\kvec,\theta,\omega)\phitilde_m(\kvec',\theta',\omega')}
    =
    2 
    \sum_{\ell=0}^\infty
    \left(A_{0}^{(2\ell)}(q)\right)^2
    \frac{\deltabar(\omega+\omega')\deltabar^2(\kvec+\kvec')}{-\imag \omega + D \kvec^2 + D_r \lambda_{2\ell}(q)/4+r}
 \ .
\end{equation}
Following the prescription in \Eref{MSD_from_GC}, this expression needs to be differentiated twice with respect to $\kvec$, before evaluating at $\kvec=\nullvec$. Most terms subsequently vanish, as
\begin{subequations}
\begin{align}
     \nabla^2|_{\kvec=\nullvec}
     \frac{k^{2j}}{-\imag\omega+Dk^2+D_r\lambda_{2\ell}(q)/4+r}
     &=\frac{v^2 d \lambda''_{2\ell}(0)/D_r - 2dD}{\left(-\imag\omega+D_r\lambda_{2\ell}(0)/4+r\right)^2} 
     \qquad\text{ for } j=0\elabel{first_deri_id}\\
     \nabla^2|_{\kvec=\nullvec}
     \frac{k^{2j}}{-\imag\omega+Dk^2+D_r\lambda_{2\ell}(q)/4+r}
     &=\frac{2d}{-\imag\omega+D_r\lambda_{2\ell}(0)/4+r} 
     \qquad\text{ for } j=1\\
     \nabla^2|_{\kvec=\nullvec}
     \frac{k^{2j}}{-\imag\omega+Dk^2+D_r\lambda_{2\ell}(q)/4+r}
     &=0  
     \qquad\text{ for } j>1 \ .
\end{align}
\end{subequations}
using $\lambda'_{2l}(q)=\plaind \lambda_{2l}(q)/\plaind q$ and $\lim_{k\to0} \lambda'_{2\ell}(q)/k= 2 \imag v \lambda''_{2\ell}(0)/D_r$ in \Eref{first_deri_id}. The MSD then derives after some algebra as
\begin{multline}\elabel{ABP_MSD}
    \spave{\xvec^2}(t) = 
    -\nabla^2_{\vec{k}=0}
    \int \dintbar{\omega}\dintbar{\omega'}\int\dTwointbar{k'} \exp{-\imag\omega t}
    \frac{1}{2\pi}
    \int_0^{2\pi} \dint{\theta'} 
    \int_0^{2\pi} \dint{\theta}
    \ave{\phi(\kvec,\theta,\omega)\phitilde_m(\kvec',\theta',\omega')}
    \\=
4Dt+2\frac{v^2}{D_r^2}\bigg[\exp{-D_r t}-1+D_r t\bigg] \ ,
\end{multline}
which is to be compared to the result for RnT in two dimensions, \Eref{MSD_dD}. Remarkably, the two are in fact identical on all time scale if the tumble rate $\tumbleRate$ is replaced by the angular diffusion constant $D_r$ \cite{Cates_2013}. As far as the MSD is concerned, $\tumbleRate^{-1}$ may be interpreted as the correlation time of the director of an RnT particle, whereas the lowest non-zero mode of the angular diffusion is $k_n=2\pi n/(2\pi)$ with $n=1$ decaying as $\exp{-k_n^2 D_r t}=\exp{-D_r t}$, so that the correlation time of the director of an active Brownian particle is indeed also $D_r^{-1}$.

\section{Discussion and Conclusion}\label{sec:discussion}

In this work, the Doi-Peliti framework \cite{Doi:1976, Peliti:1985} has been used to describe Run and Tumble (RnT) particles in arbitrary dimensions. 
We have determined the action for RnT particles in one dimension, \Erefs{def_action0_1D}, \eref{action_directly}, two \Eref{def_Apert_2D} and higher dimensions \Eref{def_Apert_dD}.
The full propagators of an RnT particle in one \Eref{full_propagator_1D}, two \Eref{full_propagator_2D} and arbitrarily higher dimensions \Eref{full_propagator_dD} have been derived, and the MSD with averaged final velocity for arbitrary dimensions \Eref{MSD_dD} has been determined. The MSD implies that the behavior of a free RnT particle is equivalent to a diffusive particle on large time scales, with an effective diffusion coefficient $D_e=D+v^2/(d\tumbleRate)$. On short time scales, however, the ballistic motion $v^2t^2$ is visible, although competing with the diffusive contribution $2Ddt$.

It is this competition that drives the rich phenomenology of RnT particles in a harmonic potential with spring-constant $k$ \cite{Garcia-MillanPruessner:2021}, as the deviation from the equilibrium behaviour vanishes with $v/(Dk)$.

For comparison, we have also determined the action \Eref{originABPaction} and subsequently the propagator for a two dimensional Active Brownian Particle (ABP) \Eref{ABP_propagator}. 
This required the use of Mathieu functions, which would probably need to be replaced by the spherical harmonics in higher dimensions, where the diffusion on the director becomes considerably more complicated, given the curved space. A perturbative approach might then become more feasible, possibly over RnT motion. The MSD of ABPs \Eref{ABP_MSD} is identical to that of RnT particles on all time scale if one identifies the tumble rate with the diffusion constant of the director of the ABPs.

In \Sref{entropy_production} we have calculated the stationary internal entropy production rate $\dot{S}=v^2/D$, \Eref{entropy_production_rate_final}, which is valid in all dimensions. In the derivation it becomes clear that one dimension has some idiosynchracies, but the entropy production in all dimensions is essentially the velocity $\vvec$ times the Stokian viscous force $\vvec k_B T/D$ acting on the particle, at temperature $T$ and with Boltzmann constant $k_B$. This is the power output of a particle that self-propels with velocity $\vvec$.

The present work provides the basis to derive characteristics of interacting RnT particles and those subject to external potentials on the basis of field theories. Using this framework provides a flexible, extensible, systmatic and perturbative approach to active matter. It comes equipped with the powerful tool of the renormalisation group and a diagrammatic that has an immediate physical interpretation, as we have attempted to illustrate above. We believe that field theory, which has been so overwhelmingly successful in the characterisation of collective phenomena \cite{WilsonandFisher:1972}, is ideally suited for the challenges that lie ahead in the research of active matter.

\section*{Acknowledgements}
The authors would like to thank Rosalba Garcia-Millan, Ignacio Bordeu and Connor Roberts for useful discussions.

\bibliographystyle{unsrt}
\end{document}